\documentclass[conference]{IEEEtran}
\IEEEoverridecommandlockouts
\usepackage{cite}
\usepackage{amsmath,amssymb,amsfonts}
\usepackage{graphicx}
\usepackage{subcaption}
\usepackage{textcomp}
\usepackage{xcolor}
\usepackage{booktabs}
\usepackage{array}
\def\BibTeX{{\rm B\kern-.05em{\sc i\kern-.025em b}\kern-.08em
    T\kern-.1667em\lower.7ex\hbox{E}\kern-.125emX}}
\begin{document}

\title{Agentic Digital Twin-Assisted Multimodal Predictive Handover}

\author{
\IEEEauthorblockN{Afan Ali, Ali Arshad Nasir, Naveed Iqbal, and Daniel Benevides da Costa}
\IEEEauthorblockA{
Research Center for Communication Systems and Sensing (IRC-CSS)\\
Department of Electrical Engineering\\
King Fahd University of Petroleum and Minerals (KFUPM)\\
Dhahran, 31261, Saudi Arabia\\
Emails: \{afan.ali@kfupm.edu.sa,~anasir@kfupm.edu.sa,~naveediqbal@kfupm.edu.sa,~danielbcosta@ieee.org\}
}
}

\maketitle
\vspace{-4mm}

\begin{abstract}
Rapid mobility, blockage, and propagation changes can invalidate instantaneous
handover decisions in dense mid-band and millimeter-wave (mmWave) networks.
We therefore propose a flexible, model-agnostic \emph{Agentic Digital Twin (DT)} framework
for predictive cell handover. The DT generates synchronized
multimodal states, while three agents autonomously select forecasts, plan
handover actions, and adapt decisions using execution feedback. The proposed Agentic DT provided the best collective balance among handover activity, the reliability of the selected links, and the accuracy of target next-generation Node B (gNB) selection relative to the best available future link. It reduced average
handovers from $11.0$ for Event~A3 to $8.0$ per trajectory while maintaining
 $98.2\%$ ray-path availability.
\end{abstract}


\vspace{-6pt}
\section{Introduction}

Reliable handover is critical in fifth-generation New Radio (5G NR) deployments, where dense microcell next-generation Node Bs (gNBs) and directional beams operate alongside legacy macrocell base transceiver stations (BTSs). Mobility, blockage, and beam-alignment changes can rapidly degrade the serving link and complicate target selection. Conventional handover relies
on the standardized 3rd Generation Partnership Project (3GPP) Event~A3
procedure with hysteresis and a time-to-trigger (TTT) interval~\cite{3gpp38331}\footnote{The hysteresis can be defined as the power margin required before initiating handover, reducing ping-pong handovers. TTT is  duration for which handover condition must remain satisfied before handover is initiated.}, and inappropriate settings can increase interruptions, radio link failures (RLFs), and ping-pong handovers. In this case, predictive mobility management and channel prediction based on large language model (LLM)
 can mitigate these limitations by anticipating future link
evolution~\cite{li2025hsr,li2026llm}. However, reliable prediction requires wireless measurements and an environmental context.

The rapid progression toward physical artificial intelligence (AI) is reshaping intelligent systems by
enabling them to perceive, reason, and act in physical environments. Digital
twins (DTs) support this progression by synchronizing physical and virtual
representations and fusing radio measurements with mobility, geometry, Light
Detection and Ranging (LiDAR), and scene information. Large language models
(LLMs), Transformer-based predictors, and agentic AI can further interpret
these multimodal states, forecast future conditions, and autonomously plan,
validate, and adapt actions~\cite{agenticDTN2026,localAgents2026}. However,
existing approaches remain insufficiently connected to site-specific handover
dynamics and event-triggered procedures in deployed networks.

\begin{figure*}[t]
  \centering
  \subfloat[]{%
    \includegraphics[width=0.9\columnwidth]{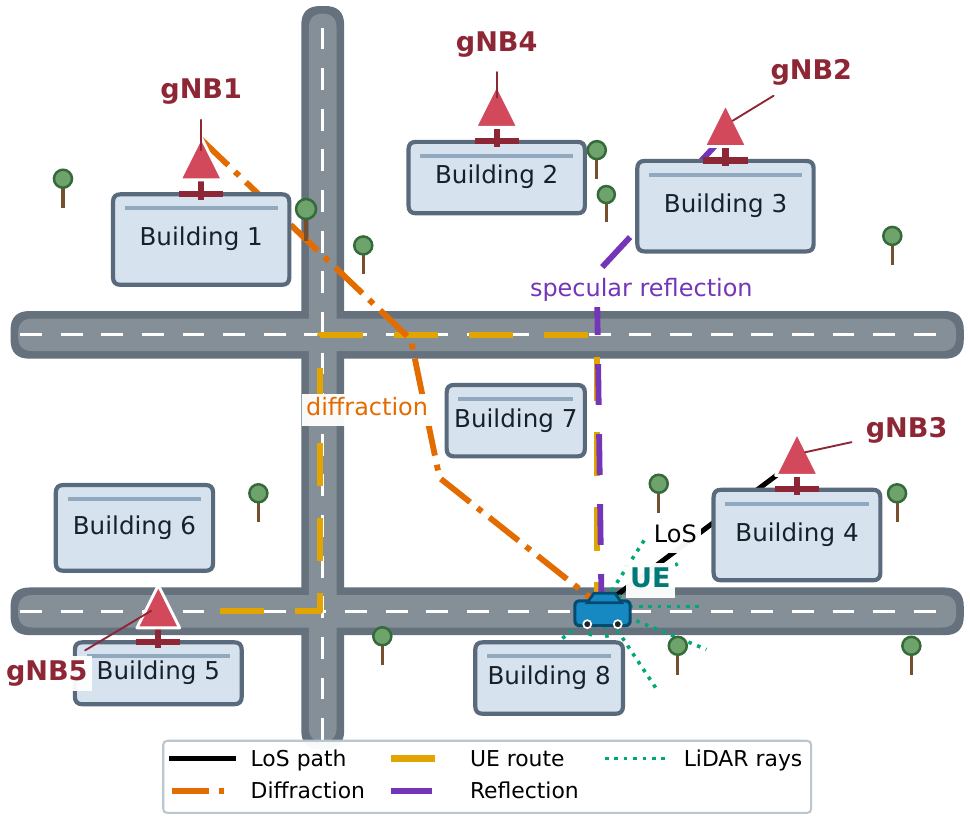}
    \label{fig:link}}
  \hfil
  \subfloat[]{%
    \includegraphics[width=0.9\columnwidth]{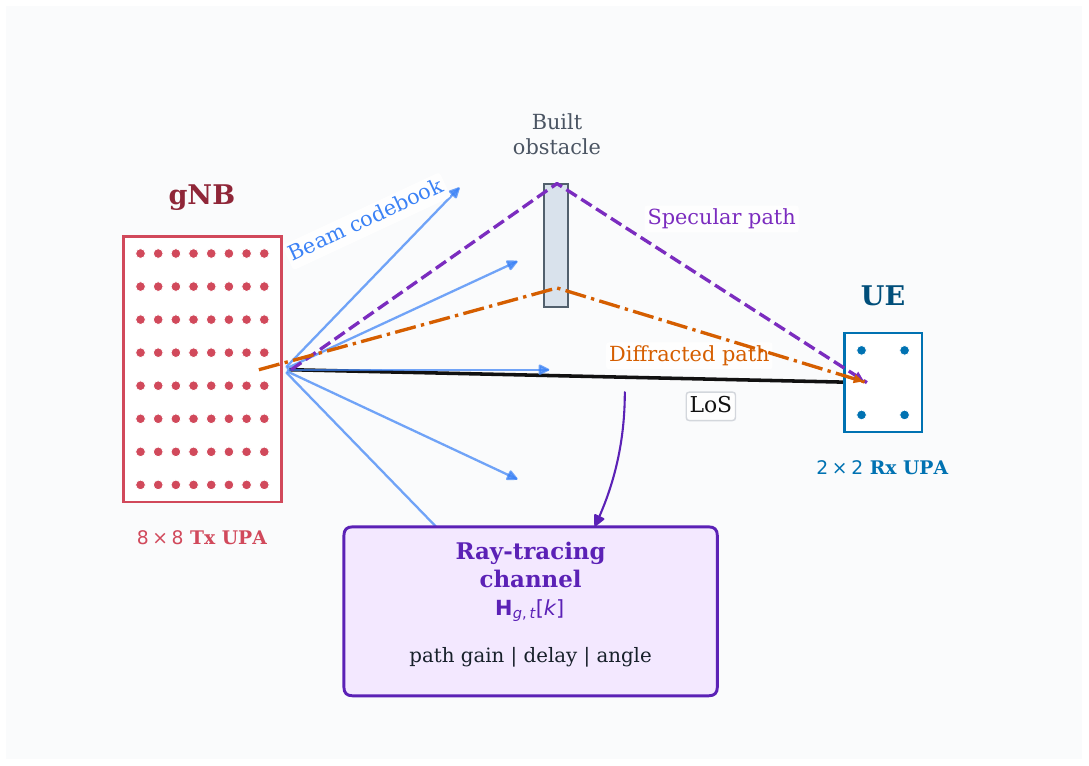}
    \label{fig:system_model}}
  \caption{Overall system model: (a)~Rooftop gNBs reaching the UE with LiDAR geometry rays along the UE route, (b)~Uniform planar array link geometry and the ray-traced channel to agentic decision pipeline.}
  \vspace{-1.5pt}
  \label{fig:over_allsystem_diagram}
\end{figure*}

Building on this progression, recent studies can be grouped into three main
directions, i.e., (i) DT-assisted predictive handover that evaluates
candidate-target suitability before execution but uses simplified channel and
sensing models~\cite{cakir2026genai}, (ii) agentic Tiny Machine Learning (ML)
for intent-aware handover on constrained devices~\cite{saleh2026agentic}, and
(iii) sensing-assisted feature-fusion methods for proactive handover and
beamforming~\cite{zhang2024mmff}. However, these approaches
address only parts of the prediction--decision--action loop. A unified
framework that predicts, selects, validates, and adapts handover actions
beyond fixed hysteresis and TTT remains insufficiently explored.

To address these gaps, we propose the \emph{Agentic DT
framework}, a flexible and model-agnostic architecture for multimodal
predictive handover. The DT generates synchronized site-specific data, while
Agentic AI selects, plans, validates, and adapts handover actions beyond fixed
hysteresis and TTT rules. This three-agent implementation represents a
preliminary step toward more autonomous handover management and
physical AI-enabled wireless systems. The main contributions can be summarized
as follows:

\begin{itemize}
\item It is developed a reproducible DT pipeline that derives virtual LiDAR and
geometric features from the site mesh and synchronizes them with ray-traced
multipath, link validity, and UE mobility. The representation is designed
to support transfer to other deployments and future integration
of live sensing and network measurements.

\item It is introduced a flexible, model-agnostic Agentic AI
framework that autonomously selects and validates handover actions. The
current implementation uses task-specific forecasting, planning, and
adaptation agents, whose policies and number can expand with site
heterogeneity, propagation clutter, computational resources, and future
LLM-based general-purpose reasoning.

\item The proposed Agentic DT is evaluated against the reactive Event~A3
baseline and non-agentic fixed policies. It provides the most balanced performance when handover activity, link validity, and target-selection alignment are considered jointly.
\end{itemize}

\section{System Model and Problem Formulation}
\label{sec:system_model}

Fig.~\ref{fig:link} depicts the scene-level propagation geometry, including rooftop gNBs, user equipment (UE) mobility, multipath, and mesh-derived LiDAR rays, while Fig.~\ref{fig:system_model} summarizes the array-level channel and data pipeline that converts ray-traced paths into synchronized radio-frequency (RF), mobility, and geometric observations for probabilistic forecasting and agentic handover.

\subsection{Network and Digital Twin Model}
\label{ssec:network_model}

The downlink mode contains $B$ candidate gNBs with overlapping coverage along the UE route, indexed by $\mathcal{B}=\{1,\ldots,B\}$. Their positions and
orientations are fixed by the site-specific DT. At route index
$t$, the mobility state $\mathbf{s}_t$ contains UE position, velocity,
acceleration, and orientation. Velocity and acceleration are derived from
consecutive route samples with interval $\Delta t$, while UE height
follows the local terrain.

The DT implements a lightweight site-specific pipeline that derives geometric
context directly from map meshes. This context contains virtual-LiDAR and
range features, depth and structural information, and line-of-sight (LoS) cues.
The mesh-based representation is portable across deployments and readily
extensible with real-time sensing and network measurements. We denote the
resulting geometric features by $\mathbf{g}_t$.

\subsection{Ray-Tracing Channel Model}
\label{ssec:channel_model}

Following the ray-tracing-based channel construction used in Sionna~\cite{aoudia2025sionnart}, the channel at gNB
$g$, route index $t$, and subcarrier $k$ can be represented as
\begin{equation}
\mathbf{H}_{g,t}[k]
=
\sum_{\ell=1}^{L_{g,t}}
\alpha_{g,\ell,t}
e^{-j2\pi f_k\tau_{g,\ell,t}}
\mathbf{a}_{\mathrm r}
\!\left(\boldsymbol{\vartheta}^{\mathrm r}_{g,\ell,t}\right)
\mathbf{a}_{\mathrm t}^{H}
\!\left(\boldsymbol{\vartheta}^{\mathrm t}_{g,\ell,t}\right),
\label{eq:rt_channel}
\end{equation}
where $\mathbf{H}_{g,t}[k]$ stand for the ray-traced channel,
$g$ indexes the candidate gNB, $t$ denotes the route index, and $k$ denotes
the subcarrier. Here, $L_{g,t}$ is the number of valid paths,
$\alpha_{g,\ell,t}$ and $\tau_{g,\ell,t}$ are the complex coefficient and
delay of path $\ell$, and $f_k$ is the frequency offset of subcarrier $k$
from the carrier frequency $f_{\mathrm c}$, giving an absolute frequency
$f_{\mathrm c}+f_k$. Considering the narrowband approximation,
$\mathbf{a}_{\mathrm r}(\cdot)$ and $\mathbf{a}_{\mathrm t}(\cdot)$ are
evaluated at $f_{\mathrm c}$, while subcarrier dependence is captured by
the delay term.

From the ray-traced path information, it is derived next two link-level RF features
for each candidate gNB: the link-validity indicator $v_{g,t}$ and the stored
path-power feature $\widetilde{P}_{g,t}$. The former identifies whether at least one
valid ray-traced propagation path exists, while the latter provides the path-based RF quantity used by the forecasting and handover modules. The path-power feature
is specific to the present ray-tracing implementation and does not represent
standardized reference signal received power (RSRP) or calibrated received power. These features can be defined
as
\begin{equation}
\begin{aligned}
v_{g,t}&=\mathbf{1}\{L_{g,t}>0\},\\
\widetilde{P}_{g,t}
&=10\log_{10}\!\left[
\max\!\left\{
\frac{1}{N_{\mathrm r}N_{\mathrm t}}
\sum_{\ell=1}^{L_{g,t}}
\left|\alpha_{g,\ell,t}\right|^2,\epsilon
\right\}\right],
\end{aligned}
\label{eq:rf_features}
\end{equation}
where $\mathbf{1}\{\cdot\}$ is the indicator function, $N_{\mathrm r}$ and
$N_{\mathrm t}$ are, respectively, the receive and transmit antenna dimensions used during
channel extraction, and $\epsilon=10^{-30}$ is the linear-power floor
($-300$~dB). For training, values are clipped at $-220$~dB and the resulting
$\widetilde{P}_{g,t}$ represents path-coefficient power in dB. Thus, the
link-validity vector across the $B$ candidate gNBs can be expressed as
\[
\boldsymbol{\ell}_t=
[v_{1,t},\ldots,v_{B,t}]^{T}.
\]
This work uses the extracted path-power and link-validity features, while full beamwise channel modeling, signal-to-interference-plus-noise ratio (SINR), and achievable-rate analysis are reserved for future beam-aware evaluation.

At each route index, the DT synchronizes UE mobility, RT radio
features, mesh-derived geometry, and link validity into the observation
vector, which can be expressed as
\begin{equation}
\mathbf{o}_t=
\left[\mathbf{s}_t,\mathbf{r}^{\mathrm{RF}}_t,
\mathbf{g}_t,\boldsymbol{\ell}_t\right],
\label{eq:multimodal_observation}
\end{equation}
where $t$ denotes the route index. Thus, $\mathbf{o}_t$ jointly captures the
current mobility, radio, geometric, and link-feasibility states, while the
separate validity vector distinguishes unavailable links from weak valid
links. The observation blocks are summarized in
Table~\ref{tab:multimodal_features}.

\begin{table}[t]
\centering
\caption{Multimodal observation blocks.}
\label{tab:multimodal_features}
\scriptsize
\begin{tabular}{@{}p{0.27\columnwidth}p{0.63\columnwidth}@{}}
\toprule
Feature block & Brief definition \\
\midrule
Mobility $\mathbf{s}_t$ &
UE position, velocity, acceleration, and orientation. \\

Radio $\mathbf{r}^{\mathrm{RF}}_t$ &
Stacked per-gNB path-power, valid-path-count, and minimum-delay features. \\

Geometry $\mathbf{g}_t$ &
Mesh-derived virtual-LiDAR, range, depth, and structural features. \\

Validity $\boldsymbol{\ell}_t$ &
Per-gNB valid-path count and derived binary validity mask. \\
\bottomrule
\end{tabular}
\end{table}

For comparison, we use a simplified Event~A3-style rule as the reactive
baseline~\cite{3gpp38331}. The complete 3GPP procedure is not modeled because
the simulation provides uncalibrated ray-traced path power rather than
standardized RSRP. Accordingly, filtered
path power can be used in the following comparison as follows
\begin{equation}
\overline{P}_{c_t,t}
-\overline{P}_{b_t^{\mathrm{serv}},t}
\geq H_{\mathrm{A3}}+O_{\mathrm{A3}},
\label{eq:a3_condition}
\end{equation}
where $b_t^{\mathrm{serv}}$ is the serving gNB, $c_t$ is the strongest
alternative, and $\overline{P}_{b,t}$ is the filtered path power. Here,
$H_{\mathrm{A3}}$ denotes the hysteresis margin and $O_{\mathrm{A3}}$ denotes
the net offset.\footnote{The net offset is a configured bias representing the
applicable cell- or measurement-specific offset in the Event~A3 comparison.}

The left-hand side therefore measures the filtered power advantage of the
strongest alternative over the serving gNB. The Event~A3 condition is
considered satisfied only when this advantage exceeds the configured margin
and remains valid for the configured TTT interval, after
which measurement reporting and handover preparation are initiated.

\vspace{-6pt}
\subsection{Handover Problem Formulation}
\label{ssec:problem_formulation}

At route index $t$, the UE is connected to serving gNB
$b_t^{\mathrm{serv}}$ and has access to the synchronized observation
$\mathbf{o}_t$. The handover problem is to decide whether to remain with the
serving gNB or prepare a transfer to another candidate before the action is
applied. It is noteworthy that an instantaneous strongest-link decision is insufficient because
mobility, blockage, and propagation changes may alter the target link during
the preparation and execution interval.

Accordingly, the set of available actions can be defined as
\begin{equation}
\mathcal{A}_t=
\{\mathrm{stay}\}\cup
\{\mathrm{HO}(b):b\in\mathcal{B}\setminus
\{b_t^{\mathrm{serv}}\}\},
\label{eq:action_set}
\end{equation}
where $\mathcal{A}_t$ is the action set, $\mathrm{stay}$ denotes retaining
the serving gNB, and $\mathrm{HO}(b)$ denotes handover to candidate gNB $b$. Thus, the controller can either maintain the current
connection or evaluate every eligible non-serving target.

Given these alternatives, the desired action is selected by maximizing
predicted future quality while accounting for handover and execution risks, which can be written as
\begin{equation}
\begin{aligned}
a_t^\star
&=\underset{a\in\mathcal{A}_t}{\operatorname{arg\,max}}
\Bigg[
\mathbb{E}\!\left(Q_{a,t:t+T_{\mathrm{exec}}}\right)
-\lambda_{\mathrm h}\mathbf{1}\{a\neq\mathrm{stay}\}\\[-1mm]
&\hspace{25mm}
-\lambda_{\mathrm f}\Pr\!\left(\mathcal{F}_{a,t}\right)
\Bigg],
\end{aligned}
\label{eq:handover_problem}
\end{equation}
where $Q_{a,t:t+T_{\mathrm{exec}}}$ is the future quality associated with
action $a$, $T_{\mathrm{exec}}$ is the decision-to-application interval,
$\lambda_{\mathrm h}$ is the handover penalty, $\lambda_{\mathrm f}$ is the execution-failure penalty and $\mathcal{F}_{a,t}$ denotes
an invalid or failed execution event. Hence, the objective rewards sustained
future quality while discouraging unnecessary handovers and unreliable
actions, i.e., its expectation and failure probability are estimated from future
channel predictions.

\section{Proposed Agentic Digital Twin-Assisted Predictive Handover Framework}
\label{sec:proposed}

\begin{figure}[t]
\centering
\includegraphics[width=0.8\linewidth]{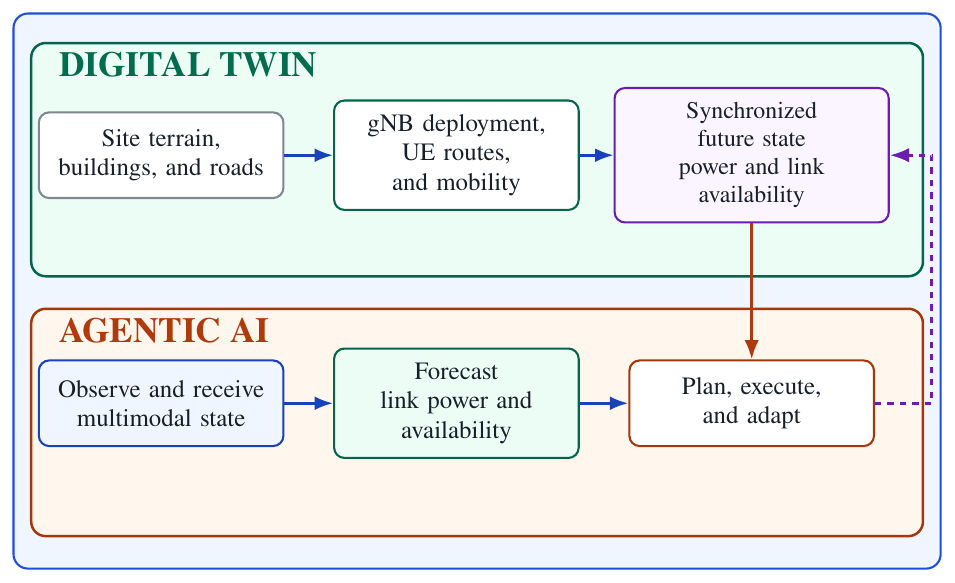}
\caption{Proposed Agentic Digital Twin-assisted framework.}
\label{fig:framework}
\end{figure}

Fig.~\ref{fig:framework} presents the proposed \emph{Agentic DT framework},
comprising a site-specific DT and an Agentic AI decision layer.
The DT generates synchronized multimodal states, while the Agentic AI autonomously forecasts, plans,
validates, and adapts handover actions using execution feedback.

\subsection{Site-Specific DT Representation and Data Generation}
\label{ssec:dt_layer}

The first component of Fig.~\ref{fig:framework} combines the site-specific
physical deployment with its virtual DT representation. In this
work, a mesh-based map models the terrain, buildings, materials, fixed gNB
locations, UE routes, antenna configurations, and propagation interactions in
the ray-tracing model of \eqref{eq:rt_channel}. The DT alos acts as
generative because it can produce site-specific radio data and
controlled ``what-if'' scenarios under different mobility, blockage, and
propagation conditions. Such scenarios enable the Agentic AI to learn and
operate under complex conditions that are difficult or impractical to
reproduce in the physical world.

For each candidate gNB and route sample, the ray-tracing solver records valid
path counts, delays, complex path coefficients, interaction information, and
link validity. The corresponding path-power and validity features are derived
using \eqref{eq:rf_features}. The observation module then synchronizes these
radio features with UE mobility and mesh-derived virtual-LiDAR and geometric
features to form the multimodal state in
\eqref{eq:multimodal_observation}. These records provide the current inputs
and reference targets for forecasting future path power and link availability.
Invalid links are retained through validity masks rather than numerical zeros.
Although the present study uses one site mesh, the representation is flexible
and deployment-agnostic.

\begin{figure}[t]
\centering
\includegraphics[width=0.5\textwidth]{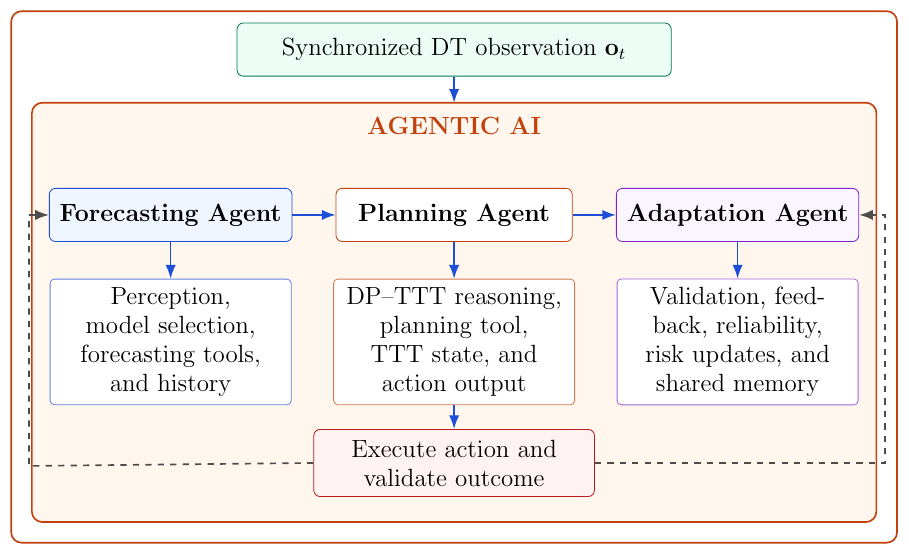}
\caption{AI agent architecture.}
\label{fig:agentic_handover}
\end{figure}

\subsection{Agentic AI Autonomous Handover}
\label{ssec:agentic_layer}

Figure~\ref{fig:agentic_handover} illustrates the architecture of agents employed. It comprises three cooperating agents: a forecasting agent, a planning agent, and an adaptation agent. The three agents work together as one flexible decision system rather than as separate neural networks. Each agent performs a specific task and can
use different models or algorithms. 

\subsubsection{Forecasting Agent}
\label{sssec:forecasting_agent}

The forecasting agent receives the synchronized observation
$\mathbf{o}_{t}$. In this work, $\mathcal{M}$ denotes the set of forecasting
candidates, comprising persistence \cite{beck2025naive}, deterministic gated
recurrent unit (GRU) \cite{helmy2023gru}, independent probabilistic GRU
\cite{song2025nqfrnn}, joint probabilistic GRU
\cite{li2025probabilistic}, and hurdle-aware joint GRU
\cite{ganjali2025hurdle} forecasters. These models serve as the agent's
internal forecasting tools. Rather than selecting a model using prediction
error alone, the agent evaluates each candidate according to its usefulness
for the handover objective in \eqref{eq:handover_problem}. The agent selects the forecaster, which can be defined as
\begin{equation}
m_t^\star
=
\underset{m\in\mathcal{M}}{\operatorname{arg\,max}}\;
S_m\!\left(\mathbf{o}_{t},\boldsymbol{\rho}_t\right),
\label{eq:forecast_agent}
\end{equation}
where $m_t^\star$ is the model selected at epoch $t$, $S_m(\cdot)$ is a
policy-aware score, and $\boldsymbol{\rho}_t$ contains the current model and
gNB reliability states. Specifically, $S_m$ is the negative exponentially
smoothed first-step masked path-power mean absolute error (MAE), computed for
all models using the validity mask. When available, predicted link
availability contributes only to the policy failure-risk estimate, otherwise,
realized validity is obtained from the same mask. Initialized from validation
error and updated causally, the score selects the model with the highest
$S_m$. The selected forecaster then supplies the future link-quality and
failure-risk estimates in \eqref{eq:handover_problem}, while the
model-agnostic design permits additional forecasters as resources and data
increase.

\subsubsection{Planning Agent}
\label{sssec:planning_agent}

The planning agent receives the selected forecaster
$m_t^\star$ and its predicted future link quality and execution risk. It uses
finite-horizon dynamic programming (DP)\footnote{DP compares possible
``stay'' and handover choices at successive future steps and selects the
sequence with the best cumulative predicted outcome.} with TTT persistence to
evaluate state-dependent action sequences. Using the forecasts from
$m_t^\star$, the maximizing sequence can be illustrated as
\begin{equation}
\begin{aligned}
a_{t:t+H_t}^{\star}
=
\underset{\substack{a_{t+h}\in\mathcal{A}_{t+h}\\
h=0,\ldots,H_t}}{\operatorname{arg\,max}}
\sum_{h=0}^{H_t}
\Big[
&\mathbb{E}\!\left(
Q^{(m_t^\star)}_{a_{t+h},t+h}
\right)
-\lambda_{\mathrm h}
\mathbf{1}\{a_{t+h}\neq\mathrm{stay}\}\\[-1mm]
&-\lambda_{\mathrm f}
\Pr\!\left(
\mathcal{F}^{(m_t^\star)}_{a_{t+h},t+h}
\right)
\Big],
\end{aligned}
\label{eq:planning_agent}
\end{equation}
where $a_{t:t+H_t}^{\star}$ is the optimal action sequence, $H_t$ is the
selected prediction horizon, and $\mathcal{A}_{t+h}$ is the feasible,
state-dependent action set in \eqref{eq:action_set}, generated by preceding
serving-cell transitions and the TTT-persistence state $c_{t+h}$. The predicted quality and failure probability include the
uncertainty and availability information supplied by the forecasting agent.
Only the first action $a_t^\star$ of the maximizing sequence is executed, and
the sequence is recomputed at the next epoch using new observations and
execution feedback.

\subsubsection{Adaptation Agent}
\label{sssec:adaptation_agent}

The adaptation agent receives the action $a_t^\star$ selected
by the planning agent, the realized link-validity sequence, and the next
observation $\mathbf{o}_{t+1}$. It uses the execution-validation function
$\mathcal{E}$ as an internal tool to determine whether the selected action
remained usable during the execution interval, which can be written as
\begin{equation}
z_t
=
\mathcal{E}\!\left(
a_t^\star,\boldsymbol{\ell}_{t:t+T_{\mathrm{exec}}}
\right),
\label{eq:execution_validation}
\end{equation}
where $z_t$ records the validity and execution outcome of the action. Using this feedback, $\mathcal{A}$ denotes the deterministic adaptation map
that updates the shared state, which can be expressed as
\begin{equation}
\boldsymbol{\rho}_{t+1}
=
\mathcal{A}\!\left(
\boldsymbol{\rho}_t,z_t,a_t^\star,\mathbf{o}_{t+1}
\right),
\label{eq:adaptation_agent}
\end{equation}
where $\boldsymbol{\rho}_t$ contains the per-gNB reliability, forecasting
weights, and planning risk state. In particular, $\mathcal{A}$ updates the selected-gNB reliability by
exponentially smoothing its previous reliability with the newly observed link
validity. It also updates the forecasting scores, failure state, and recent handover activity using the same execution outcome. Successful actions therefore reinforce reliable
gNBs and forecasting models, whereas invalid actions reduce their reliability
and increase the corresponding planning risk. The updated state is passed to
the forecasting and planning agents, allowing execution feedback to influence
subsequent model selection and action planning.

\vspace{-6pt}
\section{Simulation Setup}
\label{sec:simulation}

\subsection{Site-Specific Deployment}
\label{ssec:site_deployment}

\begin{figure}[t]
\centering
\includegraphics[width=0.8\linewidth]{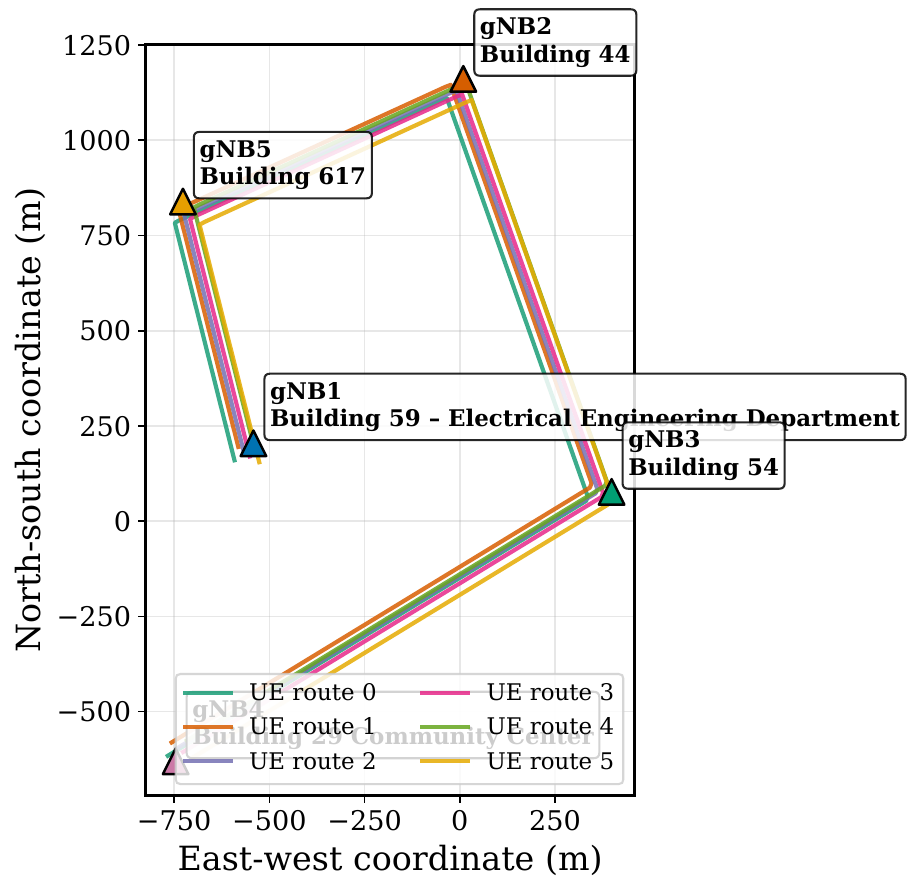}
\caption{Site-specific deployment and RF observations.}
\label{fig:updated_scene_route}
\end{figure}

In this work, the real campus of King Fahd University of Petroleum and
Minerals (KFUPM) is represented by a site-specific radio-environment DT. Its
terrain, building meshes, materials, antenna locations, and route geometry
are used to generate synchronized multimodal data for the Agentic AI layer.
The selected gNB coordinates and anchor separations are listed in
Table~\ref{tab:gnb_deployment}, while the route and radio configurations are
summarized in Table~\ref{tab:route_radio_config}. Fig.~\ref{fig:updated_scene_route}
shows six distinct, slightly shifted single-UE trajectories, each
corresponding to a separate leave-one-trajectory-out (LOTO) fold rather than
simultaneous UEs following the same path. 

All ray-tracing computations were performed using an NVIDIA GeForce RTX 5090
graphics processing unit (GPU) with Compute Unified Device Architecture
(CUDA) acceleration. An eight-point CUDA pilot was first used to verify the
Sionna RT solver and blockage behavior. The final dataset was then generated
from six routes with 240 samples each, yielding 1,440 synchronized records.

\begin{table}[t]
\centering
\caption{Five-gNB deployment on the site-specific map.}
\label{tab:gnb_deployment}
\scriptsize
\setlength{\tabcolsep}{2pt}
\begin{tabular}{@{}c l c r@{}}
\toprule
gNB & Structure & Position [m] & Dist. to gNB1 [m] \\
\midrule
gNB1 & Bldg.~59 (Start) &
$(-541.70,\,204.40,\,69.00)$ & 0.00 \\
gNB2 & Bldg.~44 &
$(9.22,\,1161.06,\,53.16)$ & 1103.95 \\
gNB3 & Bldg.~54 &
$(398.56,\,76.74,\,46.07)$ & 948.88 \\
gNB4 & Bldg.~29 Ctr. &
$(-745.97,\,-630.72,\,74.32)$ & 859.74 \\
gNB5 & Bldg.~617 &
$(-726.94,\,839.11,\,60.50)$ & 661.19 \\
\bottomrule
\end{tabular}
\end{table}
\begin{table}[t]
\centering
\caption{Route and radio configuration.}
\label{tab:route_radio_config}
\scriptsize
\setlength{\tabcolsep}{2pt}
\begin{tabular}{@{}p{0.31\columnwidth}p{0.61\columnwidth}@{}}
\toprule
Parameter & Configuration \\
\midrule
Route bank &
Six spatially offset routes; waypoint order
gNB1--gNB5--gNB2--gNB3--gNB4; 240 samples per route \\

Mobility &
5, 8, and 12~m/s labels used as covariates \\

Carrier / bandwidth / power &
3.5~GHz / 100~MHz / 44~dBm per gNB \\

Arrays &
$8\times8$ transmit and $2\times2$ receive planar arrays; 0.5-$\lambda$ spacing \\

Path solver &
Deterministic depth 5; LoS, specular reflection, and refraction enabled; diffuse reflection disabled \\

Propagation geometry &
Site-specific meshes and materials; roads and landscaping are visualization overlays only \\
\bottomrule
\end{tabular}
\end{table}

\vspace{-10pt}
\subsection{Generated Dataset and Benchmarks}
\label{ssec:data_validation}

The site-specific scene, five-gNB layout, deterministic GPU path solver, and
route bank were validated before model training. The pipeline generates
synchronized RF, mobility, and mesh-derived LiDAR records. The RF input uses
the path-based power and link-validity features defined in
\eqref{eq:rf_features}, where invalid links are retained through binary masks rather
than treated as physical zero-power observations. Since the stored RF power is
uncalibrated dB, absolute dBm thresholds are deferred to a calibrated rerun.

The dataset and forecasting targets are summarized in Table~\ref{tab:data_statistics}. The six LOTO
folds provide one held-out trajectory per fold. The best usable-gNB counts in Table~\ref{tab:data_statistics} sum to 1,431 rather than 1,440 because nine route samples contain no usable gNB, i.e, all candidate links have zero valid
ray-traced paths at those samples.

\begin{table}[t]
\centering
\caption{Generated dataset.}
\label{tab:data_statistics}
\scriptsize
\setlength{\tabcolsep}{1pt}
\begin{tabular}{@{}p{0.32\columnwidth}p{0.58\columnwidth}@{}}
\toprule
Quantity & Value / derivation \\
\midrule
Records &
1,440 records from 6 routes with 240 samples per route \\

Multimodal input &
15 RF, 10 mobility, and 72 LiDAR features; 97 features in total; mean LiDAR hit fraction 0.6561 \\

RF/link arrays &
$\texttt{rf\_power\_db}$ and
$\texttt{valid\_path\_counts}\in\mathbb{R}^{1440\times5}$ \\

Usable-path fraction &
gNB1--gNB5: 0.726, 0.758, 0.561, 0.473, and 0.793 \\

Best usable-gNB count &
gNB1--gNB5: 25, 446, 459, 240, and 261 records \\

LOTO target &
$(6,5)$ per window: six future steps for five candidate gNBs \\

Validation folds &
Six trajectory-level LOTO folds; one held-out trajectory per fold \\
\bottomrule
\end{tabular}
\end{table}

The five forecasting implementations are used as benchmarks and
are evaluated under different policy configurations. Their implementation details
and corresponding policy combinations are summarized in
Table~\ref{tab:forecast_benchmarks}. Each forecast-based
benchmark uses the same DT observations, input windows, and six-fold LOTO
protocol. The non-agentic fixed DP--TTT policy is paired with each pre-determined forecaster, while
reactive Event~A3 and standalone persistence provide additional baselines.

\begin{table}[t]
\centering
\caption{Benchmarks employed alongwith their desciption.}
\label{tab:forecast_benchmarks}
\scriptsize
\begin{tabular}{@{}p{0.27\columnwidth}p{0.37\columnwidth}p{0.27\columnwidth}@{}}
\toprule
Forecasting method & Implementation & Policy combination \\
\midrule
Persistence &
Repeats the latest observed link-power vector
\cite{beck2025naive} &
Fixed DP--TTT + persistence \\

Deterministic GRU &
Predicts future link power for each candidate gNB
\cite{helmy2023gru} &
Fixed DP--TTT + deterministic GRU \\

Independent probabilistic GRU &
Predicts separate distributions for each candidate gNB
\cite{song2025nqfrnn} &
Fixed DP--TTT + independent GRU \\

Joint probabilistic GRU &
Predicts link power jointly across candidate gNBs
\cite{li2025probabilistic} &
Fixed DP--TTT + joint GRU \\

Hurdle-aware joint GRU &
Jointly predicts link power and link availability using
continuous and binary heads \cite{ganjali2025hurdle} &
Fixed DP--TTT + hurdle-aware GRU \\

-- &
-- &
Reactive Event~A3  \\

Selected forecasting model &
Forecasting agent selects the policy-suitable model &
Proposed Agentic Twin \\
\bottomrule
\end{tabular}
\end{table}

\subsection{Results and Analysis}

\begin{figure*}[t]
\centering
\subfloat[Handover activity]{%
\includegraphics[width=0.32\textwidth]{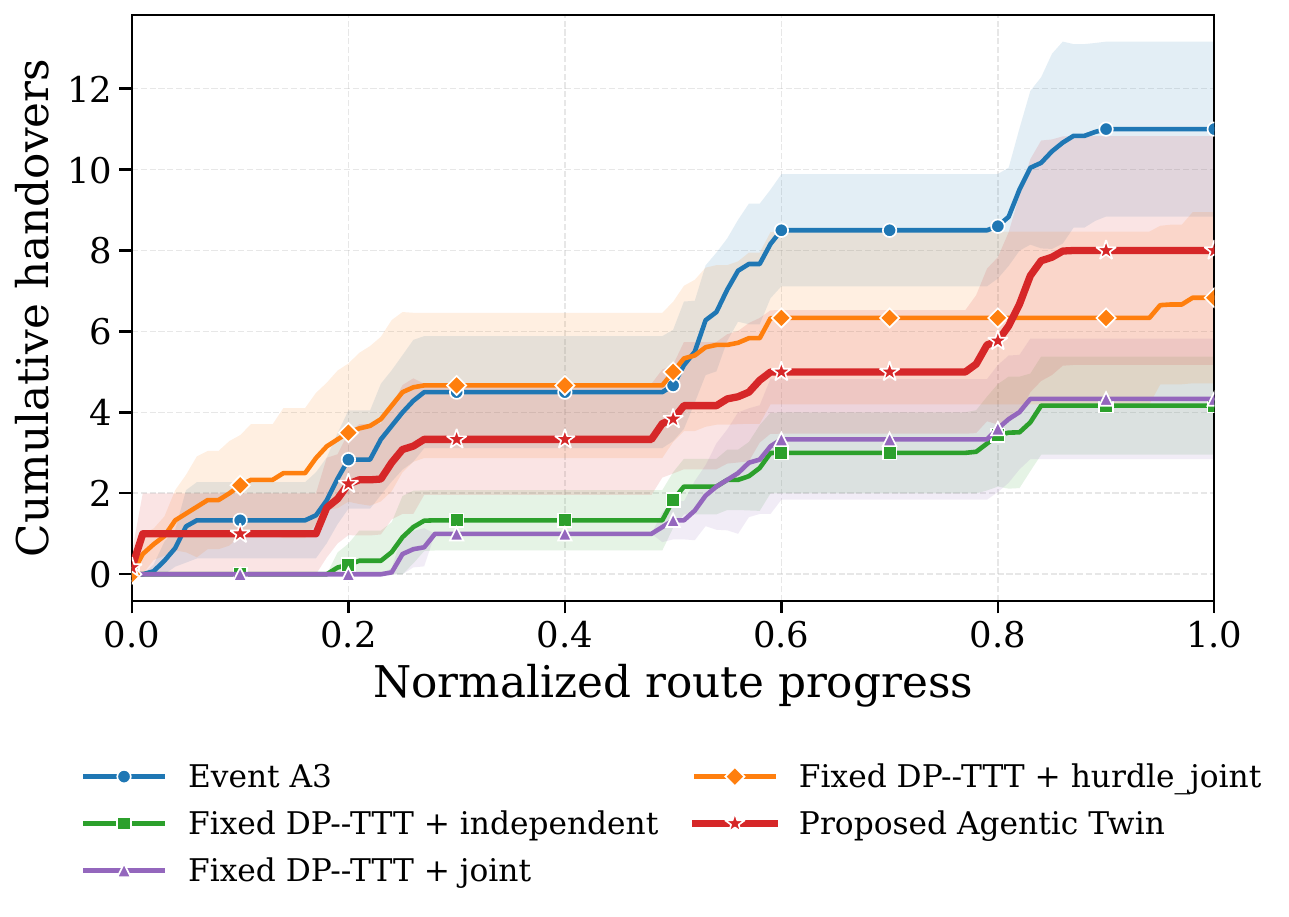}
\label{fig:handover_activity}}
\hfill
\subfloat[Link-validity behavior]{%
\includegraphics[width=0.32\textwidth]{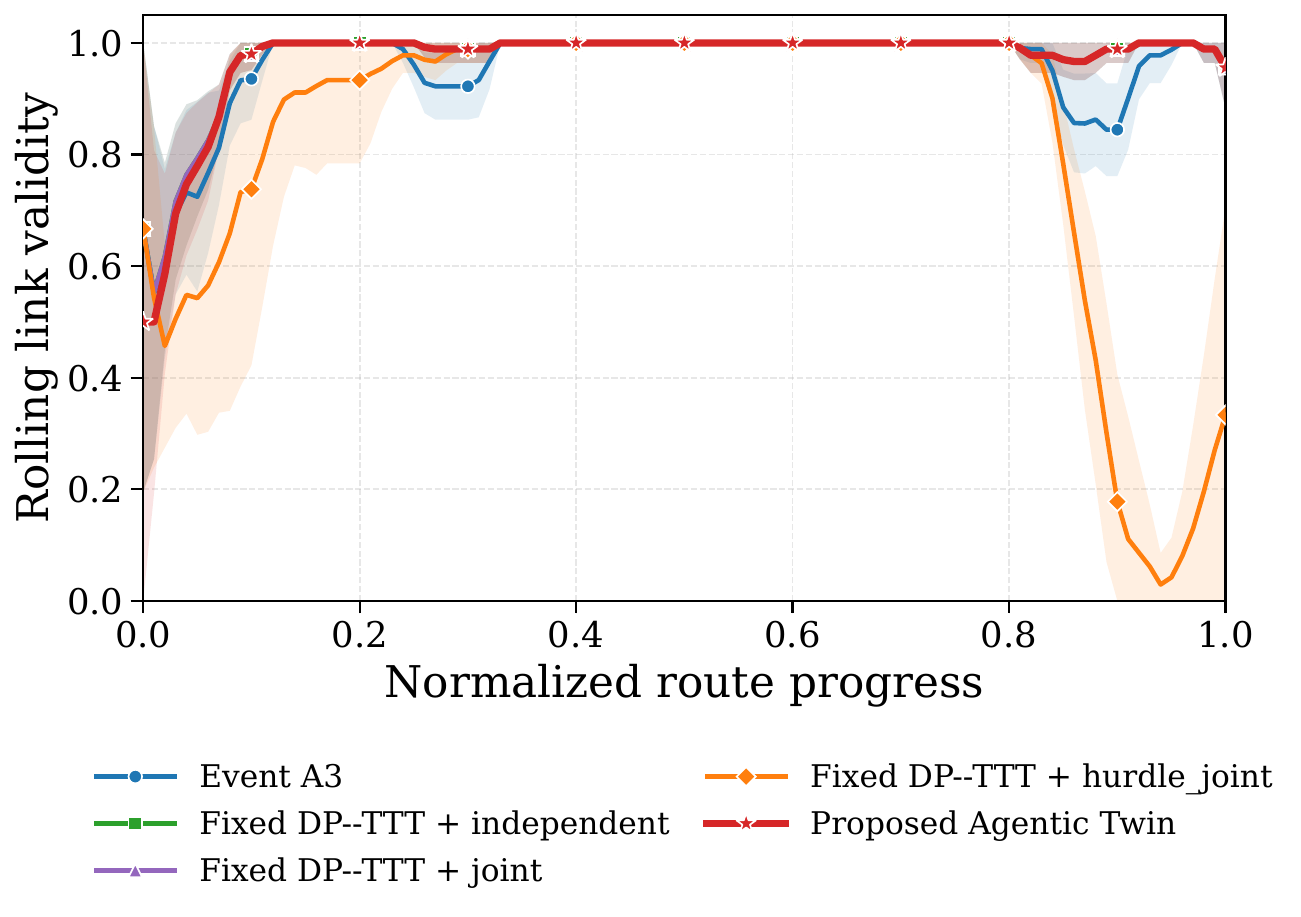}
\label{fig:link_validity}}
\hfill
\subfloat[Target-selection alignment]{%
\includegraphics[width=0.32\textwidth]{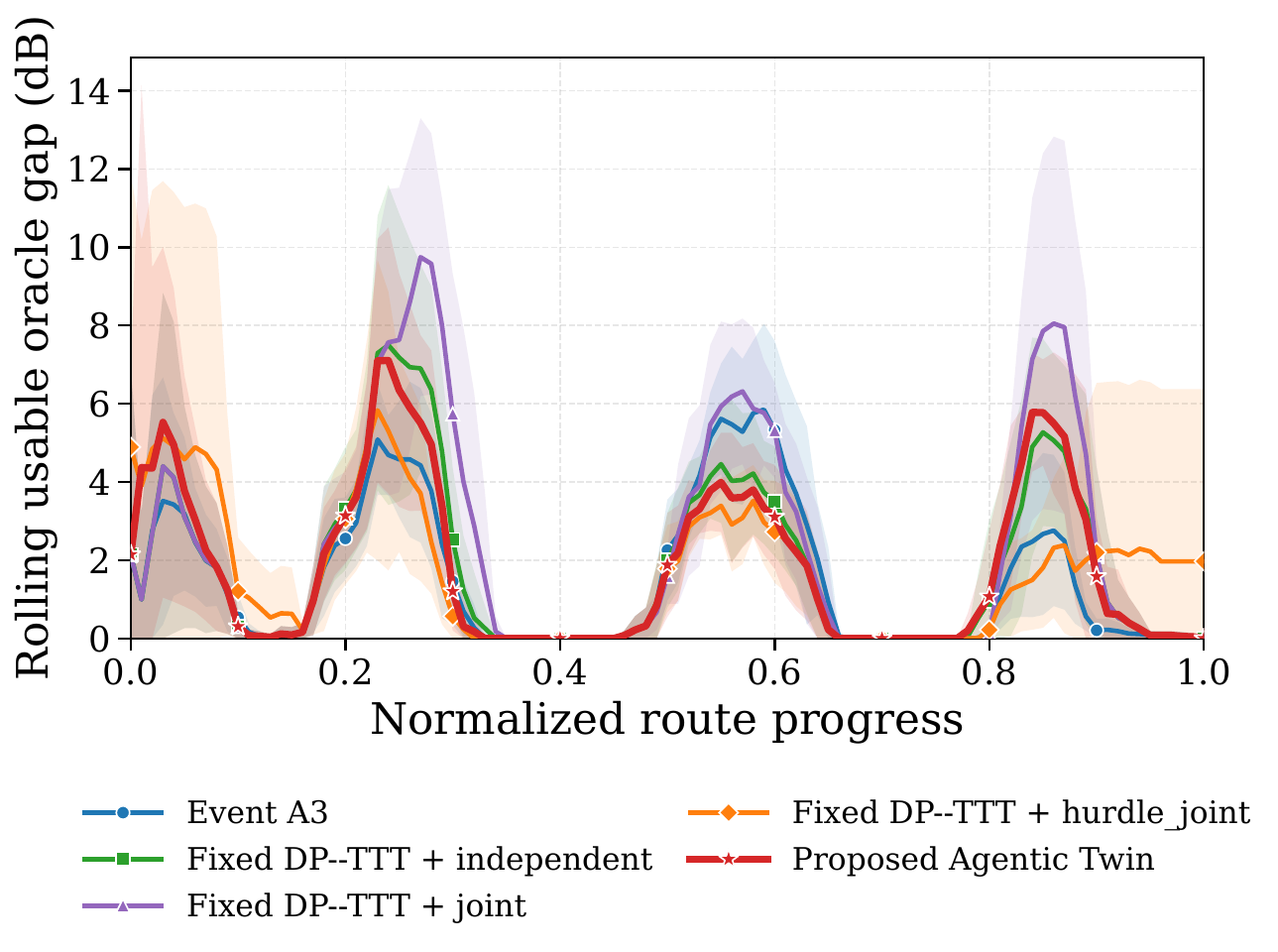}
\label{fig:target_alignment}}
\caption{Route-wise comparison of the proposed Agentic DT with other benchmarks.}
\label{fig:handover_behavior}
\end{figure*}

We evaluate each method along the normalized UE route using
handover count, link validity, and target-selection alignment. The route
progress is computed as $p_i=i/(N-1)$, where $i$ is the sample number along the
trajectory and $N$ is the total number of samples. Hence, $p=0$ and $p=1$
represent the route start and end, respectively. Fig.~\ref{fig:handover_behavior}
compares the proposed Agentic Twin with the other benchmarks.

Fig.~\ref{fig:handover_activity} plots the cumulative number of handovers
along the UE route. It shows that the proposed Agentic Twin reduces
cumulative handovers compared with Event~A3. However, the fixed independent
and joint policies produce even fewer handovers. Therefore, a lower handover
count alone does not necessarily indicate better overall performance, since
link validity and target-selection quality must also be considered.

Fig.~\ref{fig:link_validity} shows the rolling fraction of selected actions
whose target gNB has at least one valid ray-traced propagation path along the
normalized UE route. This metric measures ray-path availability rather than
successful communication or completed handover. Across the six folds, the proposed Agentic Twin achieves a mean first-step ray-path availability of approximately $98.2\%$ and maintains high availability
over most of the route, with small reductions near route transitions.

Fig.~\ref{fig:target_alignment} shows the usable oracle gap, defined as the
future-power difference between the selected target and the best valid target
at each route position. A value near zero indicates that the policy selected
the best usable gNB. The proposed Agentic Twin remains close to this oracle
over most route segments, while its occasional peaks occur during rapid
propagation changes, when target selection becomes more difficult.

\begin{figure}[t]
\centering
\includegraphics[width=1\linewidth]{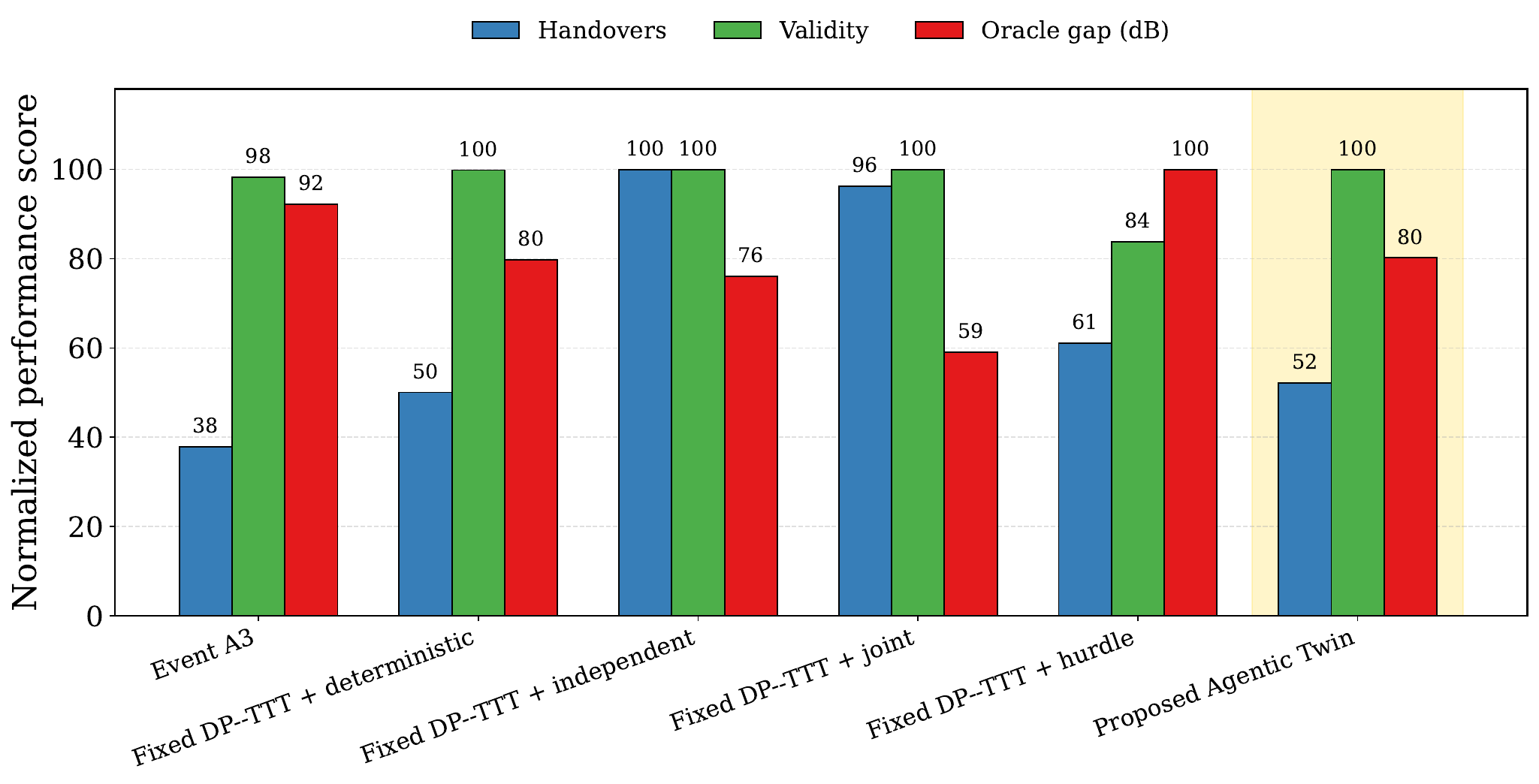}
\caption{Normalized end-to-end handover performance.}
\label{fig:overall_balance}
\end{figure}

Fig.~\ref{fig:overall_balance} summarizes end-to-end
performance using handover activity, link validity, and usable target-selection
alignment. Each score is normalized across the plotted methods. For link
validity, the score is
$100\times(\text{method mean}/\text{largest method mean})$. For handover
activity and oracle gap, where lower values are preferred, the score is
$100\times(\text{smallest method mean}/\text{method mean})$. The proposed Agentic Twin
achieves the highest validity score and remains competitive in handover
activity and oracle alignment. Although fixed policies perform better for
individual objectives, the three agents provide the most balanced performance
across all metrics.

\vspace{-6pt}
\section{Conclusion}
\label{sec:conclusion}

This work presented the flexible, model-agnostic \emph{Agentic DT framework}
for predictive handover. It combines a site-specific DT with
multimodal forecasting and three agents for planning and execution-feedback
adaptation. The proposed Agentic Twin provides a favorable trade-off among handover activity, link validity, and target-selection alignment. Future work may make the three agents more autonomous by incorporating
LLM-based general-purpose reasoning for goal
interpretation, forecasting, and planning strategy selection.


\begin{thebibliography}{00}
\footnotesize

\bibitem{3gpp38331}
3GPP, ``NR; Radio Resource Control (RRC); Protocol specification,''
3GPP TS 38.331.

\bibitem{li2025hsr}
W. Li, W. Chen, S. Wang, Y. Zhang, M. Matthaiou, and B. Ai,
``AI-driven mobility management for high-speed railway communications: Compressed measurements and proactive handover,''
\emph{IEEE Veh. Technol. Mag.}, vol. 20, no. 4, pp. 84-94, Dec. 2025.


\bibitem{li2026llm}
W. Li, W. Chen, B. Ai, Y. Zhang, Y. Peng, and M. Matthaiou,
``LLM-empowered wireless channel quality prediction for reliable cell handover,''
\emph{IEEE Wireless Commun.}, early access, May 2026.


\bibitem{agenticDTN2026}
R. Khaldi, A. Lehmann, B. Ghita, and U. Trick,
``Agentic-AI framework for integrated design, implementation, testing, and operation of DT networks,''
\emph{IEEE Open J. Commun. Soc.}, vol. 7, pp. 4352--4375, 2026.

\bibitem{localAgents2026}
G. Restuccia, D. Garlisi, F. Giuliano, A. Pagano, and I. Tinnirello,
``From Digital Twins to autonomous local agents: Rethinking AI for smart water networks,''
\emph{IEEE Internet Things J.}, vol. 13, no. 18, pp. 41534--41551, Sep. 2026.


\bibitem{cakir2026genai}
L. V. Cakir, M. A. Erturk, M. Ozdem, and B. Canberk,
``Digital Twin assisted handover scheme for mobile networks using generative AI,''
\emph{IEEE Trans. Netw. Service Manag.}, vol. 23, pp. 4562-4571, 2026.

\bibitem{saleh2026agentic}
A. Saleh, R. Morabito, S. Tarkoma, A. Lindgren, S. Pirttikangas, and L. Lov\v{e}n,
``Agentic TinyML for intent-aware handover in 6G wireless networks,''
\emph{IEEE Wireless Commun.}, Apr. 2026.


\bibitem{zhang2024mmff}
H. Zhang, S. Gao, X. Cheng, and L. Yang,
``Integrated sensing and communications toward proactive beamforming in mmWave V2I via multi-modal feature fusion,''
\emph{IEEE Trans. Wireless Commun.}, vol. 23, no. 11, pp. 15721-15735, 2024.

\bibitem{aoudia2025sionnart}
F. Ait Aoudia, J. Hoydis, M. Nimier-David, S. Cammerer, and A. Keller,
``Sionna RT: Technical report,'' NVIDIA, Tech. Rep., 2025.



\bibitem{beck2025naive}
N. Beck, J. Dovern, and S. Vogl,
``Mind the naive forecast! A rigorous evaluation of forecasting models for
time series with low predictability,''
\emph{Appl. Intell.}, vol. 55, Art. no. 395, 2025.

\bibitem{helmy2023gru}
I. Helmy, P. Tarafder, and W. Choi,
``LSTM-GRU model-based channel prediction for one-bit massive MIMO system,''
\emph{IEEE Trans. Veh. Technol.}, vol. 72, no. 8,
pp. 11053--11057, 2023.

\bibitem{song2025nqfrnn}
J. Song, W. Chang, and J. W. Song,
``NQF-RNN: Probabilistic forecasting via neural quantile function-based
recurrent neural networks,''
\emph{Appl. Intell.}, vol. 55, Art. no. 183, 2025.

\bibitem{li2025probabilistic}
Y. Li, C. H. Leung, and Q. Wu,
``Probabilistic learning of multivariate time series with temporal
irregularity,''
\emph{IEEE Trans. Knowl. Data Eng.}, vol. 37, no. 5,
pp. 2874--2887, 2025.

\bibitem{ganjali2025hurdle}
M. Ganjali, T. Baghfalaki, and N. Balakrishnan,
``A unified Bayesian approach for modeling zero-inflated count and
continuous outcomes,''
\emph{Commun. Stats.--Th. Meth.}, vol. 54, no. 24,
pp. 7669--7689, 2025.

\end{thebibliography}
\end{document}